\documentclass{birkjour_t2}
\usepackage{amsmath,amsthm,amssymb}

\theoremstyle{plain}

\theoremstyle{definition}

\theoremstyle{remark}

\renewcommand{\theenumi}{{\roman{enumi}}}

\usepackage{xcolor}
\usepackage{hyperref}
\usepackage{graphicx}

\usepackage{textcomp}
\definecolor{P@GrayBG}{gray}{.95}
\definecolor{P@GrayComment}{gray}{.40}
\RequirePackage{listings}
\lstdefinelanguage{egison}{%
  alsoletter={-},
  sensitive = true,
  comment=[l]{--},
}%
\lstdefinelanguage{haskell}{%
  sensitive = true,
  comment=[l]{--},
}%
\lstdefinelanguage{wolfram}{%
  sensitive = true,
  comment=[l]{--},
}%
\makeatletter
\newcommand\applyCurrentFontsize
{%
  \let\f@sizeS@ved\f@size%
  \let\f@baselineskipS@ved\f@baselineskip%
  \let\basicstyleS@ved\lst@basicstyle%
  \renewcommand\lst@basicstyle%
  {%
      \basicstyleS@ved%
      \fontsize{\f@sizeS@ved}{\f@baselineskipS@ved}%
      \selectfont%
  }%
}
\makeatother

\newcommand\scaledlstinline[2][]
{%
  \bgroup%
    \lstset{#1}%
    \applyCurrentFontsize%
    \lstinline|#2|%
  \egroup%
}

\usepackage{syntax}
\usepackage[symbol]{footmisc}
\usepackage{perpage}
\MakePerPage{footnote}
\usepackage{caption}
\usepackage{subcaption}
\usepackage[utf8]{inputenc}
\usepackage{pmboxdraw}

\hypersetup{breaklinks=true}
\begin{document}

\title[Symbolical Index Reduction and Completion Rules for Importing Tensor Index Notation]{Symbolical Index Reduction and Completion Rules for Importing Tensor Index Notation into Programming Languages}
\author[S. Egi]{Satoshi Egi}
\address[S. Egi]{Rakuten Institute of Technology, Rakuten, Inc., Setagayaku, Tokyo 158-0094, Japan}
\email{satoshi.egi@rakuten.com}

\maketitle

\begin{abstract}
In mathematics, many notations have been invented for the concise representation of mathematical formulae.
Tensor index notation is one of such notations and has been playing a crucial role in describing formulae in mathematical physics.
This paper shows a programming language that can deal with symbolical tensor indices by introducing a set of tensor index rules that is compatible with two types of parameters, i.e., scalar and tensor parameters.
When a tensor parameter obtains a tensor as an argument, the function treats the tensor argument as a whole.
In contrast, when a scalar parameter obtains a tensor as an argument, the function is applied to each component of the tensor.
On a language with scalar and tensor parameters, we can design a set of index reduction rules that allows users to use tensor index notation for arbitrary user-defined functions without requiring additional description.
Furthermore, we can also design index completion rules that allow users to define the operators concisely for differential forms such as the wedge product, exterior derivative, and Hodge star operator.
In our proposal, all these tensor operators are user-defined functions and can be passed as arguments of high-order functions.
\end{abstract}

\section{Introduction}

From the latter half of the twentieth century, after the first implementation of a compiler for high-level programming languages, a large amount of notations have been invented in the field of programming languages.
Lexical scoping~\cite{backus1960report}, high-order functions~\cite{mccarthy1978history,steele1996evolution}, and pattern matching~\cite{burstall1969proving,egi2018Aplas,egi2020programming} are features specific to programming invented for describing algorithms.
At the same time, researchers evolve programming languages by importing successful mathematical notations.
For example, decimal number system, function modularization, and infix notation for basic arithmetic operators have been imported in most programming languages~\cite{DBLP:journals/annals/Backus79}.
The importation of mathematical notations into programming is not always easy.
This is because the semantics of some mathematical notations are vague and complex to implement them as a part of programming languages.

This paper discusses a method for importing tensor index notation invented by Ricci and Levi-Civita~\cite{ricci1900methodes} for dealing with high-order tensors into programming languages.
Tensor calculus often appears in the various fields of computer science.
Tensor calculus is an important application of symbolic computation~\cite{DBLP:journals/pcs/KorolkovaKS13}.
Tensor calculus is heavily used in computational physics~\cite{Hughes2000} and computer visions~\cite{hartley2003multiple}.
Tensor calculus also appears in machine learning to handle multidimensional data~\cite{DBLP:journals/access/JiWLL19}.
The importation of tensor index notation makes programming in these fields easy.

The current major method for dealing with tensors is using a special syntax for describing loops for generating multi-dimensional arrays.
The \lstinline{Table}~\cite{wolframTable} expression from the Wolfram language is such a syntax construct.
$X_{ij} + Y_{ij}$ is represented as follows with the \lstinline{Table} expression.
The following program assumes that all dimensions corresponding to each index of the tensor are a constant \lstinline{M}.
\begin{lstlisting}[language=wolfram]
Table[X[[i,j]] + Y[[i,j]],{i,M},{j,M}]
\end{lstlisting}
For contracting tensors, we use the \lstinline{Sum}~\cite{wolframSum} expression inside \lstinline{Table}.
$X_{ik} Y_{kj}$ is represented as follows.
\begin{lstlisting}[language=wolfram]
Table[Sum[X[[i,k]] * Y[[k,j]],
          {k,M}],
      {i,M},{j,M}]
\end{lstlisting}
This method has the advantage that we can use an arbitrary function defined for scalar values for tensor operations.
The following Wolfram program represents ${\partial X_{ij}}/{\partial x_k}$.
\lstinline{D} is the differential function in the Wolfram language.
\begin{lstlisting}[language=wolfram]
Table[D[X[[i,j]],x[[k]]],{i,M},{j,M},{k,M}]
\end{lstlisting}
Due to this advantage, the Wolfram language has been used by mathematicians in actual research.~\cite{maeda2016geometry,maeda2010program}
However, in this method, we cannot modularize tensor operators such as tensor multiplication by functions.
Due to this restriction, we cannot syntactically distinguish applications of different tensor operators, such as tensor multiplication, wedge product, and Lie derivative, in programs.
This is because we need to represent these tensor operators combining \lstinline{Table} and \lstinline{Sum} every time when we use them.
Modularization by functions is also important for combining index notation with high-order functions.
If we can pass tensor operators to high-order functions, we can represent a formula like ``$X_{i_{1}} X_{i_{2}} ... X_{i_{n}}$'' (the number of tensors multiplied depends on the parameter $n$) by passing the operator for tensor multiplication to the fold function~\cite{DBLP:books/daglib/0067889}.
There are other existing works that take the same approach with the Wolfram language.
NumPy's \lstinline{einsum} operator~\cite{numpy}, Diderot's EIN operator~\cite{kindlmann2016diderot}, and tensor comprehensions~\cite{DBLP:journals/corr/abs-1802-04730} are such work.
Some of them provide a syntactic construct whose appearance is similar to mathematical formulae.
However, they have the same restriction on function modularization.
This restriction comes from the requirement that users need to specify the indices of the result tensors (e.g., ``$_{ij}$'' of $A_{ij} = X_{ik} Y_{kj}$) for determining whether to contract a pair of identical symbolic indices.

Maxima~\cite{maximaWeb,toth2005tensor,solomonik2015sparse} takes a different approach from Wolfram.
In Maxima, we describe formulae of tensor calculus using several special operators prepared for tensors, such as $+$ and $\cdot$, that support tensor index notation.
$+$ is a function that sums the components of two tensors given as arguments.
$\cdot$ is a function for tensor multiplication.
It takes the tensor product of the two tensors given as arguments and takes the sum of the trace if there are pairs of a superscript and a subscript with the same index variable.
Using this method enables index notation to be directly represented in a program as the mathematical expressions.
However, in this method, index notation can be used only by functions that are specially prepared to use it.
One of the reasons is that index rules differ for each operator.
For example, $+$ and $\cdot$ work in the different way for the same arguments such as $A_i + B_i$ and $A_i \cdot B_i$.
$A_i + B_i$ returns a vector, but $A_i \cdot B_i$ returns a scalar.
Ahalander's work~\cite{aahlander2002einstein} that implements index notation on C++ takes the same approach.

Array-oriented programming languages, such as APL and J, take a completely different approach.
They do not use tensor index notation for representing tensor calculus.
Instead, they invented a new notion, \emph{function rank}~\cite{bernecky1987introduction}.
Function rank specifies how to map the operators to the components of tensors.
When the specified function rank is $0$ for an argument matrix, the operator is mapped to each scalar component of the matrix ($A_i + B_{jk}$).
When the specified function rank is $1$ for an argument matrix, the operator is mapped to the rows of the matrix by regarding the matrix as a vector of vectors ($A_j + B_{ij}$).
When the specified function rank is $2$ for an argument matrix, the operator is applied to the matrix directly ($A_i + B_{ij}$).
\begin{lstlisting}[language=wolfram]
J> (2 $ 1 2) +"1 0 (2 2 $ 10 20  30 40)
11 12
21 22

31 32
41 42
J> (2 $ 1 2) +"1 1 (2 2 $ 10 20  30 40)
11 22
31 42
J> (2 $ 1 2) +"1 2 (2 2 $ 10 20  30 40)
11 21
32 42
\end{lstlisting}
A similar idea to the function rank is also imported into various programming languages and frameworks, including Wolfram~\cite{wolframListable} and NumPy~\cite{scipyBroadcasting}.
However, the function rank has a limitation that it does not allow to represent an expression that requires transposition of an argument tensor: e.g., $A_{ij} + B_{ji}$ (this expression requires the transposition of the matrix $B$).

This paper shows that a combination of a set of symbolical index reduction and completion rules with scalar and tensor parameters, which is a simplified notion of function rank, enables us to use tensor index notation for arbitrary functions defined just for scalar values without requiring any additional descriptions.
In our method, we do not distinguish tensor operators that handle symbolical tensor indices with the other user-defined functions, therefore we can pass these tensor operators as arguments of high-order functions.

\section{Language Design for Importing Tensor Index Notation}

This section presents a new method for importing tensor index notation into programming languages.
Briefly, it is achieved by introducing two types of parameters, \textit{scalar parameters} and \textit{tensor parameters}, and simple index reduction rules.
First, we introduce scalar and tensor parameters.
Second, we introduce a set of index reduction rules that is compatible with them.
The combination of scalar and tensor parameters and the proposed index reduction rules enables us to apply user-defined functions to tensors using tensor index notation.
Third, we introudce index completion rules for omitted tensor indices.
By designing the index completion rules for omitted indices properly, we become able to concisely define the operators even for the differential forms~\cite{schutz1980geometrical}, such as the wedge product, exterior derivative, and Hodge star operator.
The method proposed in this paper has already been implemented in the Egison programming language~\cite{egison}.
Egison has a similar syntax to the Haskell programming language.

\subsection{Scalar and Tensor Parameters}

Scalar and tensor parameters are a similar notion to the function rank~\cite{bernecky1987introduction}.
When a scalar parameter obtains a tensor as an argument, the function is applied to each component of the tensor.
In contrast, when a tensor parameter obtains a tensor as an argument, the function treats the tensor argument as a whole.
We call a function that takes only scalar parameters \emph{scalar function} and a function that takes only tensor parameters \emph{tensor function}.
For example, ``\lstinline{+}'', ``\lstinline{-}'', ``\lstinline{*}'', and ``\lstinline{/}'' should be defined as scalar functions;
a function for multiplying tensors and a function for matrix determinant should be defined as tensor functions.
The difference between function rank and these two types of parameters is that function rank allows users to control the level of mapping, whereas scalar and tensor parameters only allow users to specify whether we map the function to each component of the tensor or not.
Instead, the proposed tensor index reduction rules combined with scalar and tensor parameters allow users to control the level of mapping.

Figure~\ref{fig:minFunction} shows the definition of the \lstinline{min} function as an example of a scalar function.
The \lstinline{min} function takes two numbers as the arguments and returns the smaller one.
``\lstinline{$}'' is prepended to the beginning of the parameters of the \lstinline{min} function.
It means the parameters of the \lstinline{min} function are scalar parameters.
When a scalar parameter obtains a tensor as an argument, the function is applied to each component of the tensor like tensor product as Figures~\ref{fig:minDiff} and~\ref{fig:minSame}.
As Figure~\ref{fig:minSame}, if the indices of the tensors given as arguments are identical, the result matrix is reduced to the vector that consists of the diagonal components.
This reduction is defined in the proposed tensor index reduction rules that will be explained in the next section.
Thus the \lstinline{min} function can handle tensors even though it is defined without considering tensors.

\begin{figure}[t]
 \begin{subfigure}[b]{1.0\linewidth}
\begin{lstlisting}[language=egison]
def min $x $y := if x < y then x else y
\end{lstlisting}
  \caption{Definition of the \lstinline{min} function}
  \label{fig:minFunction}
 \end{subfigure}
  \medskip
 \begin{subfigure}[b]{1.0\linewidth}
  {\small
    $min(\begin{pmatrix} 1 \\ 2 \\ 3 \\ \end{pmatrix}_{i},  \begin{pmatrix} 10 \\ 20 \\ 30 \\ \end{pmatrix}_{j}) = \begin{pmatrix} min(1,10) & min(1,20) & min(1,30) \\ min(2,10) & min(2,20) & min(2,30) \\ min(3,10) & min(3,20) & min(3,30) \\ \end{pmatrix}_{ij}
    = \begin{pmatrix} 1 & 1 & 1 \\ 2 & 2 & 2 \\ 3 & 3 & 3 \\ \end{pmatrix}_{ij}$
    }
  \caption{Application of the \lstinline{min} function to the vectors with different indices}
  \label{fig:minDiff}
 \end{subfigure}
  \medskip
 \begin{subfigure}[b]{1.0\linewidth}
  {\small
    $min(\begin{pmatrix} 1 \\ 2 \\ 3 \\ \end{pmatrix}_{i},  \begin{pmatrix} 10 \\ 20 \\ 30 \\ \end{pmatrix}_{i}) = \begin{pmatrix} min(1,10) & min(1,20) & min(1,30) \\ min(2,10) & min(2,20) & min(2,30) \\ min(3,10) & min(3,20) & min(3,30) \\ \end{pmatrix}_{ii}
  = \begin{pmatrix} min(1,10) \\ min(2,20) \\ min(3,30) \end{pmatrix}_{i} = \begin{pmatrix} 1 \\ 2 \\ 3 \end{pmatrix}_{i}$ 
    }
  \caption{Application of the \lstinline{min} function to the vectors with identical indices}
  \label{fig:minSame}
 \end{subfigure}
 \caption{Definition and application of \lstinline{min} function}
  \label{fig:scalarFunction}
\end{figure}

Figure~\ref{fig:dotFunction} shows the definition of the ``\lstinline{.}'' function as an example of a tensor function.
``\lstinline{.}'' is a function for multiplying tensors.
``\lstinline{
It means the parameters of the ``\lstinline{.}'' function are tensor parameters.
As with ordinary functions, when a tensor is provided to a tensor parameter, the function treats the tensor argument as a whole maintaining its indices.
``\lstinline{.}'' is defined as an infix operator.
For defining an infix operator, we enclose the name of a function with parenthesis.
In Figure~\ref{fig:dotFunction}, ``\lstinline{+}'' and ``\lstinline{*}'' are scalar functions for addition and multiplication, respectively.
\lstinline{contract} is a primitive function to contract a tensor that has pairs of a superscript and subscript with identical symbols.
Figure~\ref{fig:dotSame} shows the example for calculating the inner product of two vectors using the ``\lstinline{.}'' function.
We can use the ``\lstinline{.}'' function for any kind of tensor multiplication such as tensor product and matrix multiplication as well as inner product.

\begin{figure}[t]
 \begin{subfigure}[b]{1.0\linewidth}
\begin{lstlisting}[language=egison]
def (.) %t1 %t2 := contract (+) (t1 * t2)
\end{lstlisting}
  \caption{Definition of the ``\lstinline{.}'' function}
  \label{fig:dotFunction}
 \end{subfigure}
  \medskip
 \begin{subfigure}[b]{1.0\linewidth}
  {\small
  $\begin{pmatrix} 1 \\ 2 \\ 3 \\ \end{pmatrix}^{i} \cdot \begin{pmatrix} 10 \\ 20 \\ 30 \\ \end{pmatrix}_{i} = contract(+, \begin{pmatrix} 10 & 20 & 30 \\ 20 & 40 & 60 \\ 30 & 60 & 90 \\ \end{pmatrix}^{i}_{\;i})
  = 10 + 40 + 90 = 140$
  \\ \medskip \\
  $\begin{pmatrix} 1 \\ 2 \\ 3 \\ \end{pmatrix}_{i} \cdot \begin{pmatrix} 10 \\ 20 \\ 30 \\ \end{pmatrix}_{i} = contract(+, \begin{pmatrix} 10 \\ 40 \\ 90 \\ \end{pmatrix}_{i}) = \begin{pmatrix} 10 \\ 40 \\ 90 \\ \end{pmatrix}_{i}$
  \\ \medskip \\
  $\begin{pmatrix} 1 \\ 2 \\ 3 \\ \end{pmatrix}_{i} \cdot \begin{pmatrix} 10 \\ 20 \\ 30 \\ \end{pmatrix}_{j} = contract(+, \begin{pmatrix} 10 & 20 & 30 \\ 20 & 40 & 60 \\ 30 & 60 & 90 \\ \end{pmatrix}_{ij})
  = \begin{pmatrix} 10 & 20 & 30 \\ 20 & 40 & 60 \\ 30 & 60 & 90 \\ \end{pmatrix}_{ij}$ \\
  }
  \caption{Application of the ``\lstinline{.}'' function}
  \label{fig:dotSame}
 \end{subfigure}
 \caption{Definition and application of ``\lstinline{.}'' function}
\end{figure}

\begin{figure}[t]
 \begin{subfigure}[b]{1.0\linewidth}
\[R^{i}_{\;jkl} = \frac{\partial \Gamma^{i}_{\;jl}}{\partial x^k} - \frac{\partial \Gamma^{i}_{\;jk}}{\partial x^l} + \Gamma^{m}_{\;jl} \Gamma^{i}_{\;mk} - \Gamma^{m}_{\;jk} \Gamma^{i}_{\;ml} \]
  \caption{Formula of Riemann curvature tensor}
  \label{fig:inMathShort}
 \end{subfigure}
  \medskip
 \begin{subfigure}[b]{1.0\linewidth}
\begin{lstlisting}[language=wolfram]
R=Table[D[`$\Gamma$`[[i,j,l]],x[[k]]] - D[`$\Gamma$`[[i,j,k]],x[[l]]]
       +Sum[`$\Gamma$`[[m,j,l]] `$\Gamma$`[[i,m,k]]
          - `$\Gamma$`[[m,j,k]] `$\Gamma$`[[i,m,l]],
            {m,M}],
        {i,M},{j,M},{k,M},{l,M}]
\end{lstlisting}
  \caption{Wolfram program that represents the formula in Figure~\ref{fig:inMathShort}}
  \label{fig:inWolframShort}
   \end{subfigure}
  \medskip
 \begin{subfigure}[b]{1.0\linewidth}
\begin{lstlisting}[language=egison]
def R~i_j_k_l := withSymbols [m]
  `$\partial$/$\partial$` `$\Gamma$`~i_j_l x~k - `$\partial$/$\partial$` `$\Gamma$`~i_j_k x~l + `$\Gamma$`~m_j_l . `$\Gamma$`~i_m_k - `$\Gamma$`~m_j_k . `$\Gamma$`~i_m_l
\end{lstlisting}
  \caption{A program by the proposed language that represents the formula in Figure~\ref{fig:inMathShort}}
  \label{fig:inEgisonShort}
   \end{subfigure}
 \caption{Representations of Riemann curvature tensor formula}
\end{figure}

Let us show another example in differential geometry.
When the mathematical expression in Fugure~\ref{fig:inMathShort} is expressed in a standard way in the Wolfram language, it becomes a program such as the one shown in Figure~\ref{fig:inWolframShort}.
The same expression can be expressed in our system as shown in Figure~\ref{fig:inEgisonShort}.
Our system supports two types of indices, both superscripts and subscripts.
A subscript is represented by ``\lstinline|_|''.
A superscript is represented by ``\lstinline|~|''.
A double loop consisting of the \lstinline{Table} and \lstinline{Sum} expressions appears in the program in the Wolfram language, whereas the program in our system is flat, similar to the mathematical expression.
This is achieved by using tensor index notation in the program.
In particular, the reason that the loop structure by the \lstinline{Sum} expression in the Wolfram language does not appear in our expression to express $\Gamma^{m}_{\;jk} \Gamma^{i}_{\;ml} - \Gamma^{m}_{\;jl} \Gamma^{i}_{\;mk}$ is that the ``\lstinline{.}'' function can handle Einstein summation notation.
It is emphisized that the program \lstinline[mathescape]|($\partial$/$\partial$ $\Gamma$~i_j_k x~l)| in this example expresses $\frac{\partial \Gamma^{i}_{\;jk}}{\partial x^l}$ in the first term on the right-hand side.
In the Wolfram language, the differential function \lstinline{D} is applied to each tensor component, but our differential function \lstinline[mathescape]!$\partial$/$\partial$! is applied directly to the tensors.

The differential function \lstinline[mathescape]!$\partial$/$\partial$! is defined in a program as a scalar function like the \lstinline{min} function.
When a tensor is provided as an argument to a scalar function, the function is applied automatically to each component of the tensor.
Therefore, when defining a scalar function, it is sufficient to consider only a scalar as its argument.
That is, in the definition of the \lstinline[mathescape]!$\partial$/$\partial$! function, the programmers need write the program only for the case in which the argument is a scalar value.
Despite that, the program \lstinline[mathescape]|$\partial$/$\partial$ $\Gamma$~i_j_k x~l| returns a fourth-order tensor.
Thus, we can import tensor index notation including Einstein notation into programming if we clearly distinguish between tensor functions such as ``\lstinline{.}'' and scalar functions such as ``\lstinline{+}'' and \lstinline[mathescape]!$\partial$/$\partial$!.

\subsection{Reduction Rules for Tensors with Indices}

This section presents a whole set of index reduction rules that are compatible with the scalar and tensor parameters.
Let us consider the reduction rules only for a single tensor, and that is enough to define the set of the index reduction rules.
This is because the reduction rules are applied only after a scalar function is applied to tensor arguments as seen in Fig.~\ref{fig:scalarFunction}.

\subsubsection{Multiple Identical Symbolical Indices}

Tensors are reduced only when they have identical symbols as their indices.
First, let us discuss the cases that tensors have identical superscripts or subscripts.

When multiple indices of the same symbol appear, our system converts it to the tensor composed of diagonal components for these indices.
After this conversion, the leftmost index symbol remains.
For example, the indices ``\lstinline|_i_j_i|'' convert to ``\lstinline|_i_j|''.
In our system, unbound variables are treated as symbols.
These symbols can be used as indices of tensors.
In our system, a tensor is expressed by enclosing its components with ``\lstinline{[|}'' and ``\lstinline{|]}''.
We express a higher-order tensor by nesting this notation, as we do for an $n$-dimensional array.
In this paper, we show the evaluation result of a program in the comment that follows the program.
``\lstinline{--}'' is the inline comment delimiter of our language.
\begin{lstlisting}[language=egison]
[|[|11,12,13|],[|21,22,23|],[|31,32,33|]|]_i_j -- [|[|11,12,13|],[|21,22,23|],[|31,32,33|]|]_i_j

[|[|11,12,13|],[|21,22,23|],[|31,32,33|]|]_i_i -- [|11,22,33|]_i

[|[|[|1,2|],[|3,4|]|],[|[|5,6|],[|7,8|]|]|]_i_j_i -- [|[|1,3|],[|6,8|]|]_i_j
\end{lstlisting}
When three or more subscripts of the same symbol appear, our system converts it to the tensor composed of diagonal components for all these indices.
\begin{lstlisting}[language=egison]
[|[|[|1,2|],[|3,4|]|],[|[|5,6|],[|7,8|]|]|]_i_i_i -- [|1,8|]_i
\end{lstlisting}
Superscripts and subscripts behave symmetrically.
When only superscripts are used, they behave in exactly the same manner as when only subscripts are used.
\begin{lstlisting}[language=egison]
[|[|[|1,2|],[|3,4|]|],[|[|5,6|],[|7,8|]|]|]~i~j~i -- [|[|1,3|],[|6,8|]|]~i~j
\end{lstlisting}

\subsubsection{Superscripts and Subscripts with Identical Symbols}

Next, let us consider a case in which the same symbols are used for a superscript and a subscript.
In this case, some existing work~\cite{aahlander2002einstein} automatically contracts the tensor using ``$+$''.
In contrast, our system just converts it to the tensor composed of diagonal components, as in the above case.
However, in this case, the summarized indices become a \textit{supersubscript}, which is represented by ``\lstinline|~_|''.
\begin{lstlisting}[language=egison]
[|[|11,12,13|],[|21,22,23|],[|31,32,33|]|]~i_i -- [|11,22,33|]~_i
\end{lstlisting}
Even when three or more indices of the same symbol appear that contain both superscripts and subscripts, our system converts it to the tensor composed of diagonal components for all these indices.
\begin{lstlisting}[language=egison]
[|[|[|1,2|],[|3,4|]|],[|[|5,6|],[|7,8|]|]|]~i~i_i -- [|1,8|]~_i
\end{lstlisting}
The reason not to contract it immediately is to parameterize an operator for contraction.
The components of supersubscripts can be contracted by using the \lstinline{contract} expression.
The \lstinline{contract} expression receives a function to be used for contraction as the first argument, and a target tensor as the second argument.
This feature allows us to implement a tensor multiplication function.
\begin{lstlisting}[language=egison]
contract (+) [|11,22,33|]~_i -- 66
\end{lstlisting}
In the above program, the ``\lstinline{+}'' function is passed to \lstinline{contract} as a prefix operator.
When an infix operator is enclosed with parenthesis, it becomes an prefix operator.

\subsubsection{Pseudo Code for Index Reduction}

\begin{figure}
  \begin{center}
\begin{lstlisting}[language=egison]
E({A, xs}) =
  if e(xs) = [] then
    {A, xs})
  elsif e(xs) = [{k,j}, `$\dotsc$`] & p(k,xs) = p(j,xs) then
    E({diag(k, j, A), remove(j, xs))
  elsif e(xs) = [{k,j}, `$\dotsc$`] & p(k,xs) != p(j,xs) then
    E({diag(k, j, A), update(k, 0, remove(j, xs)))
\end{lstlisting}
  \end{center}
  \caption{Pseudo code of index reduction}
  \label{fig:semanticsRules}
\end{figure}

Figure~\ref{fig:semanticsRules} shows the pseudo-code of index reduction explained in the above.
\lstinline{E(A,xs)} is a function for reducing a tensor with indices, where \lstinline{A} is an array that consists of tensor components and \lstinline{xs} is a list of indices appended to \lstinline{A}.
For example, \lstinline{E(A,xs)} works as follows with the tensor whose indices are ``\lstinline|~i_j_i|''.
We use \lstinline{1}, \lstinline{-1}, and \lstinline{0} to represent a superscript, subscript, and supersubscript, respectively.
\begin{lstlisting}[language=egison]
E({[|[|[|1,2|],[|3,4|]|]
    ,[|[|5,6|],[|7,8|]|]|]
 ,[{i,1}, {j,-1}, {i,-1}]}) =
{[|[|1,3|],[|6,8|]|], [{i,0}, {j,-1}]}
\end{lstlisting}
Let us explain the helper functions used in Figure~\ref{fig:semanticsRules}:
\lstinline{e(xs)} is a function for finding pairs of identical indices from \lstinline{xs};
\lstinline{diag(k, j, A)} is a function for creating the tensor that consists of diagonal components of \lstinline{A} for the \lstinline{k}-th and \lstinline{j}-th order;
\lstinline{remove(k, xs)} is a function for removing the \lstinline{k}-th element from \lstinline{xs};
\lstinline{p(k, xs)} is a function for obtaining the value of the \lstinline{k}-th element of the assoc list \lstinline{xs};
\lstinline{update(k, p, xs)} is a function for updating the value of the \lstinline{k}-th element of the assoc list \lstinline{xs} to \lstinline{p}.
These functions work as follows.
\begin{lstlisting}[language=egison]
e([{i, 1}, {j, -1}, {i, -1}])        = [{1,3}]
diag(1, 2, [|[|11,12|],[|21,22|]|]) = [|11,22|]
p(2, [{i, 1}, {j, -1}])             = -1
remove(2, [{i, 1}, {j, -1}])        = [{i, 1}]
update(2, 0, [{i, 1}, {j, -1}])     = [{i, 1}, {j, 0}]
\end{lstlisting}

\subsection{Implementation of Scalar and Tensor Parameters}

Let us explain how to implement scalar and tensor parameters.
The implementation of tensor parameters is the same with the ordinary parameters because the function treats the argument tensor as it is.
In contrast, a function with scalar parameters is converted to a function only with tensor parameters by using the \lstinline{tensorMap} function as follows.
\begin{lstlisting}[language=egison]
\ $x $y -> ...
-- => \ %x %y ->
--      tensorMap (\ %x -> tensorMap (\ %y -> ...) y))
--                x
\end{lstlisting}
As the name implies, the \lstinline{tensorMap} function applies the function of the first argument to each component of the tensor provided as the second argument.
When the result of applying the function of the first argument to each component of the tensor provided as the second argument is a tensor with indices, it moves those indices to the end of the tensor that is the result of evaluating the \lstinline{tensorMap} function.

Let us review the \lstinline{min} function defined in Figure~\ref{fig:minFunction} as an example.
This \lstinline{min} function can handle tensors as arguments as follows.
\begin{lstlisting}[language=egison]
min [|1,2,3|]_i [|10,20,30|]_j -- [|[|1,1,1|],[|2,2,2|],[|3,3,3|]|]_i_j
min [|1,2,3|]_i [|10,20,30|]_i -- [|1,2,3|]_i
\end{lstlisting}
Note that the tensor indices of the first evaluated result are ``\lstinline|_i_j|''.
If the \lstinline{tensorMap} function simply applies the function to each component of the tensor, the result of this program is \lstinline![|[|1 1 1|]_j [|2 2 2|]_j [|3 3 3|]_j|]_i!.
However, as explained above, if the results of applying the function to each component of the tensor are tensors with indices, it moves those indices to the end of the tensor that is the result of evaluating the \lstinline{tensorMap} function.
This is the reason that the indices of the evaluated result are ``\lstinline|_i_j|''.
This mechanism enables us to directly apply scalar functions to tensor arguments using index notation as the above example.

Next, let us review the ``\lstinline{.}'' function defined in Figure~\ref{fig:dotFunction} as an example of a tensor function.
When a tensor with indices is given as an argument of a tensor function, it is passed to the tensor function maintaining its indices.
It allows us to directly apply tensor functions to tensor arguments using index notation as in the following example.

\begin{lstlisting}[language=egison]
[|1,2,3|]~i . [|10,20,30|]_i -- 140
[|1,2,3|]_i . [|10,20,30|]_j -- [|[|10,20,30|],[|20,40,60|],[|30,60,90|]|]_i_j
[|1,2,3|]_i . [|10,20,30|]_i -- [|10,40,90|]_i
\end{lstlisting}

Tensor parameters are rarely used compared with scalar paramters because a tensor parameter is used only when defining a function that contracts tensors.

\subsection{Inverted Scalar Arguments}

The \lstinline[mathescape]!$\partial$/$\partial$! function in Figure~\ref{fig:inEgisonShort} is a scalar function.
However, \lstinline[mathescape]!$\partial$/$\partial$! is not a normal scalar function.
\lstinline[mathescape]!$\partial$/$\partial$!  is a scalar function that inverts indices of the tensor given as its second argument.
For example, the program \lstinline[mathescape]|($\partial$/$\partial$ $\Gamma$~i_j_k x~l)| returns the fourth-order tensor with the indices ``\lstinline|~i_j_k_l|''.

To define scalar functions such as \lstinline[mathescape]!$\partial$/$\partial$!, we use \textit{inverted scalar parameters}.
Inverted scalar parameters are represented by ``\lstinline{*$}''.
A program that uses inverted scalar parameters is transformed as follows.
The \lstinline{flipIndices} function is a primitive function for inverting the indices of a tensor provided as an argument upside down.
Supersubscripts remain as supersubscripts even if they are inverted.
\begin{lstlisting}[language=egison]
def `$\partial$`/`$\partial$` $f *$x := ...
-- => def `\color{P@GrayComment}{$\partial$}`/`\color{P@GrayComment}{$\partial$}` %f %x := tensorMap (\ %f -> tensorMap (\ %x -> ...) (flipIndices x))
--                            f
\end{lstlisting}
In the following example, we apply \lstinline[mathescape]!$\partial$/$\partial$! to tensors.
\begin{lstlisting}[language=egison]
`$\partial$`/`$\partial$` [|(r * (sin `$\theta$`)),(r * (cos `$\theta$`))|]_i [|r,`$\theta$`|]_j
-- [|[|(sin `\color{P@GrayComment}{$\theta$}`),(r * (cos `\color{P@GrayComment}{$\theta$}`))|],[|(cos `\color{P@GrayComment}{$\theta$}`),(-1 * r * (sin `\color{P@GrayComment}{$\theta$}`))|]|]_i~j
`$\partial$`/`$\partial$` [|(r * (sin `$\theta$`)),(r * (cos `$\theta$`))|]_i [|r,`$\theta$`|]_i
-- [|(sin `\color{P@GrayComment}{$\theta$}`),(-1 * r * (sin `\color{P@GrayComment}{$\theta$}`))|]~_i
\end{lstlisting}

\subsection{The \lstinline{withSymbols} Expression}\label{withSymbols}

The \lstinline{withSymbols} expression is syntax for generating new local symbols as the \lstinline{Module}~\cite{wolframModule} expression in the Wolfram language.
One-character symbols that are often used as indices of tensors such as \lstinline{i}, \lstinline{j}, and \lstinline{k} are often used in another part of a program.
Generating local symbols using \lstinline{withSymbols} expressions enables us to avoid variable conflicts for such symbols.

The \lstinline{withSymbols} expression takes a list of symbols as its first argument.
These symbols are valid only in the expression given in the second argument of the \lstinline{withSymbols} expression.
\begin{lstlisting}[language=egison]
withSymbols [i] contract (+) ([|1,2,3|]~i * [|10,20,30|]_i) -- 60
\end{lstlisting}
It acts in a special way when the evaluation result of the body of the \lstinline{withSymbols} expression contains the symbols generated by the \lstinline{withSymbols} expressions.
In that case, the result tensor is transposed to shift those symbols backward and remove them.
In the following evaluation result, the matrix is transposed because \lstinline{j} is shifted backward before it removed.
This mechanism is useful to handle differential forms that will be discussed in the next section.

\begin{lstlisting}[language=egison]
withSymbols [j] [|[|1,2|],[|3,4|]|]_j_i) -- [|[|1,3|],[|2,4|]|]_i
\end{lstlisting}

\subsection{Index Completion Rules for Tensors with Omitted Indices}

By designing the index completion rules for omitted indices properly, we can extend the proposed method explained so far to express a calculation handling the differential forms~\cite{schutz1980geometrical}.

There are several prior studies~\cite{karczmarczuk1999functional,sussman2013functional} for programming differential forms.
These studies prepare special data structures for differential forms and the operators for differential forms, such as exterior derivative and Hodge star operator, are implemented on these data structures.
On the other hand, our approach represents a differential form just as a multi-dimentional array as we do for tensors.
This approach allows us to define the operators for differential forms using tensor index notations and the operators defined for tensors.

\subsubsection{Differential Forms}\label{egisonPara}

In mathematics, we sometimes denote a $p$-form by appending only $(n-p)$ indices to an $n$-th order tensor.
For example, a third order tensor \lstinline[mathescape]|$\omega$~i_j| denotes a matrix-valued $1$-form.
We import this notation for $p$-forms into programming by designing the proper index completion rules that are suitable for this notation.

For general scalar functions, it is natural to complement the same indices to each argument tensor as follows.
Let \lstinline{A} and \lstinline{B} in the following examples be scalar-valued 2-forms.
\begin{lstlisting}[language=egison]
A + B -- => A_t1_t2 + B_t1_t2
\end{lstlisting}
In most of the cases, if we complement indices as above, we can represent the operation for differential forms.
However, this completion is not suitable for functions specially defined for differential forms such as the wedge product and exterior derivative.
In the case of the wedge product, we would like to append the different indices to each argument as follows.
\begin{lstlisting}[language=egison]
A `$\wedge$` B -- => A_t1_t2 `\color{P@GrayComment}{$\wedge$}` B_t3_t4
\end{lstlisting}

We introduce the ``\lstinline{!}'' operator for this purpose.
If the ``\lstinline{!}'' operator is prepended to function application, the omitted indices are complemented by the latter method.
With this method, the function for calculating the wedge product is defined in one line as follows.
\begin{lstlisting}[language=egison]
def (`$\wedge$`) %A %B := A !. B
\end{lstlisting}
We can also define the exterior derivative in one line as follows.
The \lstinline{flip} function is a function for swapping the arguments of a two-argument function.
It is used to transpose the result.
\begin{lstlisting}[language=egison]
def d %A := !(flip `$\partial$`/`$\partial$`) params A
\end{lstlisting}
Next, we can define Hodge star operator.
\begin{lstlisting}[language=egison]
def hodge %A :=
  let k := dfOrder A in
    withSymbols [i, j]
      (sqrt (abs (det g_#_#))) * (foldl (.) ((`$\epsilon$` N k)_(i_1)..._(i_N) . A..._(j_1)..._(j_k))
                                              (map (\t -> g~(i_t)~(j_t)) [1..k]))
\end{lstlisting}
It is a direct translation of the following mathematical formula for Hodge star operator~\cite{minHodge}.
\begin{equation}\label{hodge}
*A = \sqrt{\det|g|} \cdot \epsilon_{i_{1}...i_{n}} \cdot A_{j_{1}...j_{k}} \cdot g^{i_{1}j_{1}} \cdots g^{i_{k}j_{k}} \cdot e^{i_{k+1}} \wedge ... \wedge e^{i_{n}}
\end{equation}
In the above program, \lstinline{dfOrder} is a function that returns $p$ when it obtains an $p$-form;
\lstinline{det} is a function for calculating the determinant of the argument matrix;
\lstinline[mathescape]{$\epsilon$} is the Levi-Civita symbols;
``\lstinline{#}'' used in the indices as \lstinline|g_#_#| represents a dummy symbol.
All instances of ``\lstinline{#}'' are treated as different symbols.
In a program that deals with high-order tensors, the number of symbols used for indices increases.
A dummy symbol suppresses that.
This mechanism makes it easier to distinguish indices, thereby also improves the readability of the program.
Thus, we can concisely define the operators for differential forms by controlling the completion of omitted indices.
We can see the sample programs that use the functions defined above in Egison Mathematics Notebook\cite{egisonMath}.

\section{Demonstration}

This section presents a program for calculating the Riemann curvature tensor~\cite{fleisch2011student,schutz1980geometrical,ollivier2011visual}, in which the fourth-order tensor is necessary to be calculated.
Figure~\ref{fig:inEgison} is a program for calculating the Riemann curvature tensor of $S^2$ using the formula of curvature form.
Our language has an interface for Jupyter Notebook~\cite{kluyver2016jupyter}, which renders a evaluation result as mathematical formula.

In our system, when binding a tensor to a variable, we can specify the type of indices in the variable name.
For example, we can bind different tensors to \lstinline|$g__|, \lstinline|$g~~|, \lstinline[mathescape]!$\Gamma$___!, and \lstinline[mathescape]!$\Gamma$~__!.
This feature is also implemented in Maxima~\cite{maximaWeb} and simplifies variable names.
In Figure~\ref{fig:inEgison}, some of the tensors are bound to a variable with symbolical indices such as \lstinline[mathescape]!$\Gamma$_i_j_k!.
It is automatically desugared as follows.
This syntactic sugar renders a program closer to the mathematical expression.
\lstinline{transpose} is a function for transposing the tensor in the second argument as specified in the first argument.
\begin{lstlisting}[language=egison]
def `$\Gamma$`_i_j_k := ...
-- => def `\color{P@GrayComment}{$\Gamma$}`___ := withSymbols [i,j,k]
--      (transpose [i,j,k] ...)
\end{lstlisting}
The \lstinline{antisymmetrize} function is used in the definition of the curvature form \lstinline[mathescape]|$\Omega$~i_j|.
It is a function for normalizing a differential form to an anti-symmetric tensor.
It is defined in our system as a tensor function.
The \lstinline{evenAndOddPermutations} function returns a tuple of even and odd permutations of the given size.
The \lstinline{subrefs} function appends the elements of the second argument collection as the subscripts to the first argument tensor.
\begin{lstlisting}[language=egison]
def antisymmetrize %X :=
  let p := dfOrder X
      (es, os) := even-and-odd-permutations p in
    withSymbols [i]
      (sum (map (\ `$\sigma$` -> subrefs X (map (\t -> i_(`$\sigma$` t)) [1..p]) es) -
       sum (map (\ `$\sigma$` -> subrefs X (map (\t -> i_(`$\sigma$` t)) [1..p]) os)) / (fact p)
\end{lstlisting}
Figure~\ref{fig:inMath} shows the mathematical formulae of Christoffel symbols and curvature form.
These formulae appear in the 4th, 5th, and 8th programs in Figure~\ref{fig:inEgison}.

\begin{figure}[h]
 \begin{subfigure}[b]{1.0\linewidth}
  \centering
  \includegraphics[width=15.0cm,bb=0 0 1774 1530]{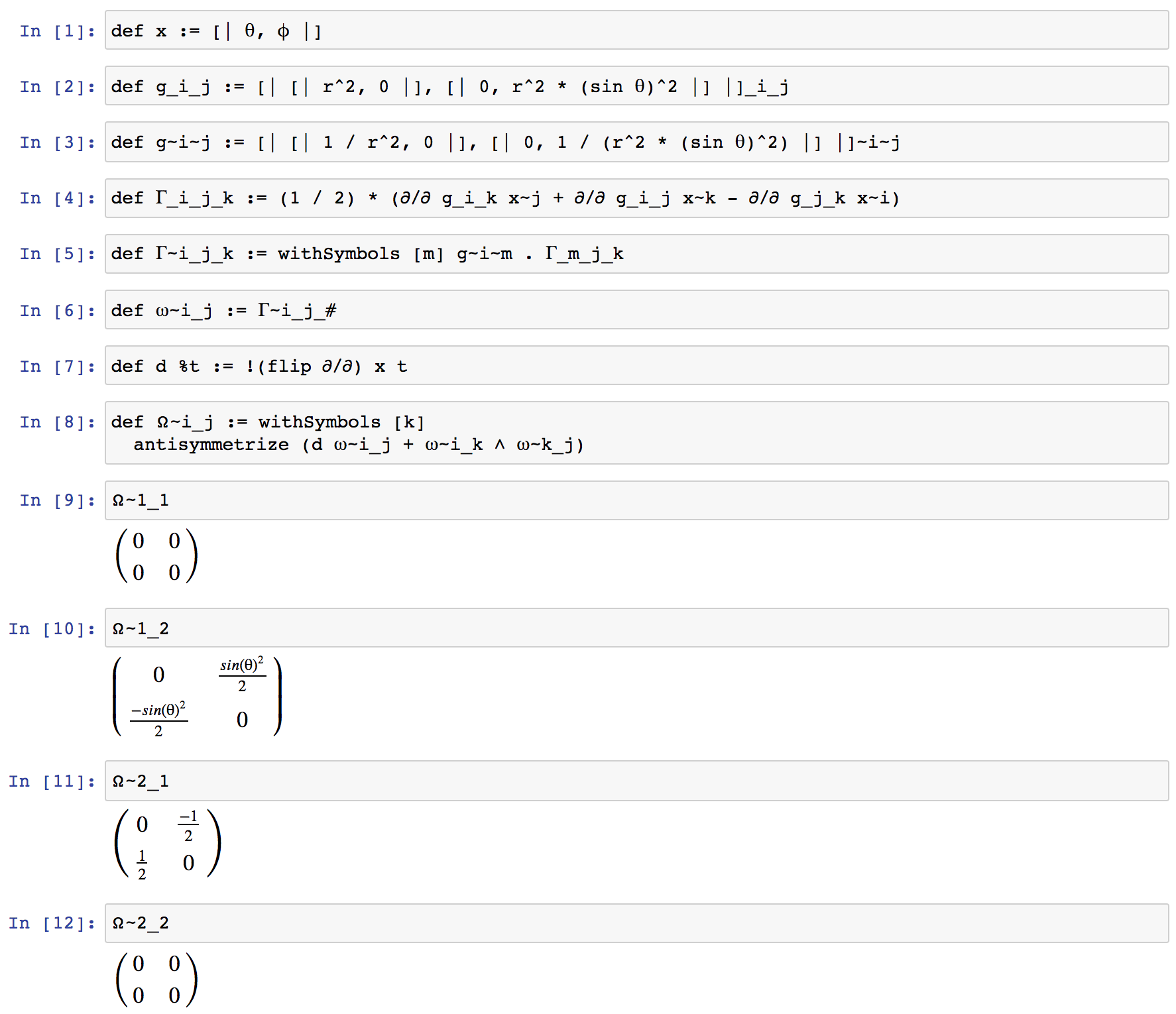}
  \caption{A program for calculating Riemann curvature tensor of $S^2$ using the formula of curvature form}
  \label{fig:inEgison}
 \end{subfigure}
  \medskip
 \begin{subfigure}[b]{1.0\linewidth}
\begin{alignat*}{3}
\Gamma_{ijk} = \frac{1}{2} (\frac{\partial g_{ij}}{\partial x^k} + \frac{\partial g_{ik}}{\partial x^j} - \frac{\partial g_{kj}}{\partial x^i}) &\qquad
\Gamma^{i}_{\;jk} = g^{im} \Gamma_{mjk} &\qquad
\Omega^{i}_{\;j} = d \omega^{i}_{\;j} + \omega^{i}_{\;k} \wedge \omega^{k}_{\;j}
\end{alignat*}
  \caption{Mathematical formulae used in the above program}
  \label{fig:inMath}
 \end{subfigure}
 \caption{Program for differential geometry in our language}
\end{figure}

\section{Conclusion}

This paper showed that we can import tensor index notation into a programming language by introducing a set of symbolical index reduction rules that is compatible with scalar and tensor parameters.
The proposed method allows users to define functions that handle symbolic tensor indices without additional descriptions for handling these symbolic tensor indices.
We demonstrated the proposed method by showing our proof-of-concept programming language Egison.
The proposed method can be implemented also in the other programming languages.
It is an interesting research topic to think about how to incorporate the proposed method into existing programming languages.
For example, we think it is possible to introduce the concept of scalar and tensor parameters using a static type system~\cite{DBLP:books/daglib/0005958}.
In a language with a static type system, whether the parameter of a function is a scalar or tensor parameter can be specified when specifying the type of the function.

In particular, it is of substantial significance to incorporate the proposed method into programming languages such as Formura~\cite{muranushi2016automatic} and Diderot~\cite{kindlmann2016diderot} that have a compiler to generate code for executing tensor calculation efficiently.
For example, incorporating the proposed method into Formura would enable us to describe physical simulation using not only the Cartesian coordinate system but also more general coordinate systems such as the polar and spherical coordinate systems in simple programs.

This paper focuses on the importation of tensor index notation into programming.
However, there are still many notations in mathematics that are useful but not yet introduced into programming.
Importation of these notations is not only useful but also leads us to deepen our understanding on mathematical notations.

\section*{Acknowledgements}
The author thanks Hiroki Fukagawa for motivating the author to import the notation for differential forms into programming languages.
The author thanks Yutaka Shikano for intensive discussions on the content of the paper.
The author thanks Momoko Hattori and Mayuko Kori for developing a user-friendly interface of the proposed system.
The author thanks Hiromi Hirano, Hidehiko Masuhara, and Michal J. Gajda for helpful feedback on the earlier versions of the paper.

\bibliographystyle{abbrv}
\bibliography{main}

\begin{thebibliography}{10}

\bibitem{wolframListable}
{Listable - Wolfram Language Documentation}.
\newblock \url{http://reference.wolfram.com/language/ref/Listable.html}, 2020.

\bibitem{maximaWeb}
{Maxima - a Computer Algebra System}.
\newblock \url{http://maxima.sourceforge.net/}, 2020.

\bibitem{wolframModule}
{Module - Wolfram Language Documentation}.
\newblock \url{http://reference.wolfram.com/language/ref/Module.html}, 2020.

\bibitem{numpy}
{NumPy}.
\newblock \url{http://www.numpy.org/}, 2020.

\bibitem{wolframSum}
{Sum - Wolfram Language Documentation}.
\newblock \url{http://reference.wolfram.com/language/ref/Sum.html}, 2020.

\bibitem{wolframTable}
{Table - Wolfram Language Documentation}.
\newblock \url{http://reference.wolfram.com/language/ref/Table.html}, 2020.

\bibitem{scipyBroadcasting}
{Universal functions (ufunc) --- NumPy v1.16 Manual}.
\newblock \url{https://docs.scipy.org/doc/numpy/reference/ufuncs.html}, 2020.

\bibitem{aahlander2002einstein}
K.~{\AA}hlander.
\newblock Einstein summation for multidimensional arrays.
\newblock {\em Computers \& Mathematics with Applications}, 44(8-9):1007--1017,
  2002.

\bibitem{DBLP:journals/annals/Backus79}
J.~W. Backus.
\newblock The history of {FORTRAN} i, {II} and {III}.
\newblock {\em {IEEE} Annals of the History of Computing}, 1(1):21--37, 1979.

\bibitem{backus1960report}
J.~W. Backus, F.~L. Bauer, J.~Green, C.~Katz, J.~McCarthy, P.~Naur, A.~Perlis,
  H.~Rutishauser, K.~Samelson, B.~Vauquois, et~al.
\newblock {Report on the algorithmic language ALGOL 60}.
\newblock {\em Numerische Mathematik}, 2(1):106--136, 1960.

\bibitem{bernecky1987introduction}
R.~Bernecky.
\newblock An introduction to function rank.
\newblock {\em ACM SIGAPL APL Quote Quad}, 18(2):39--43, 1987.

\bibitem{DBLP:books/daglib/0067889}
R.~S. Bird and P.~Wadler.
\newblock {\em Introduction to functional programming}.
\newblock Prentice Hall International series in computer science. Prentice
  Hall, 1988.

\bibitem{burstall1969proving}
R.~M. Burstall.
\newblock Proving properties of programs by structural induction.
\newblock {\em The Computer Journal}, 12(1):41--48, 1969.

\bibitem{egisonMath}
S.~Egi.
\newblock Egison mathematics notebook.
\newblock \url{http://www.egison.org/math/}, 2017.

\bibitem{egison}
S.~Egi.
\newblock {The Egison Programming Language}.
\newblock \url{http://www.egison.org/}, 2020.

\bibitem{egi2018Aplas}
S.~Egi and Y.~Nishiwaki.
\newblock Non-linear pattern matching with backtracking for non-free data
  types.
\newblock In {\em Asian Symposium on Programming Languages and Systems (APLAS
  2018)}, pages 3--23. Springer, 2018.

\bibitem{egi2020programming}
S.~Egi and Y.~Nishiwaki.
\newblock Functional programming in pattern-match-oriented programming style.
\newblock In {\em The Art, Science, and Engineering of Programming}, 2020.

\bibitem{fleisch2011student}
D.~A. Fleisch.
\newblock {\em A student's guide to vectors and tensors}.
\newblock Cambridge University Press, 2011.

\bibitem{hartley2003multiple}
R.~Hartley and A.~Zisserman.
\newblock {\em Multiple view geometry in computer vision}.
\newblock Cambridge university press, 2003.

\bibitem{Hughes2000}
T.~J.~R. Hughes.
\newblock {\em {The Finite Element Method: Linear Static and Dynamic Finite
  Element Analysis}}.
\newblock Dover Civil and Mechanical Engineering. Dover, 2000.

\bibitem{DBLP:journals/access/JiWLL19}
Y.~Ji, Q.~Wang, X.~Li, and J.~Liu.
\newblock {A Survey on Tensor Techniques and Applications in Machine Learning}.
\newblock {\em {IEEE} Access}, 7:162950--162990, 2019.

\bibitem{karczmarczuk1999functional}
J.~Karczmarczuk.
\newblock Functional coding of differential forms.
\newblock In {\em Scottish Workshop on Functional Programming}, 1999.

\bibitem{kindlmann2016diderot}
G.~Kindlmann et~al.
\newblock Diderot: a domain-specific language for portable parallel scientific
  visualization and image analysis.
\newblock {\em IEEE transactions on visualization and computer graphics},
  22(1):867--876, 2016.

\bibitem{kluyver2016jupyter}
T.~Kluyver, B.~Ragan-Kelley, F.~P{\'e}rez, B.~E. Granger, M.~Bussonnier,
  J.~Frederic, K.~Kelley, J.~B. Hamrick, J.~Grout, S.~Corlay, et~al.
\newblock Jupyter notebooks - a publishing format for reproducible
  computational workflows.
\newblock In {\em ELPUB}, pages 87--90, 2016.

\bibitem{DBLP:journals/pcs/KorolkovaKS13}
A.~V. Korolkova, D.~S. Kulyabov, and L.~A. Sevastyanov.
\newblock {Tensor computations in computer algebra systems}.
\newblock {\em Programming and Computer Software}, 39(3):135--142, 2013.

\bibitem{maeda2010program}
Y.~Maeda, S.~Rosenberg, and F.~Torres-Ardila.
\newblock {Computation of the Wodzicki-Chern-Simons form in local coordinates.
  Computations for $S^1$ actions on $S^2 \times S^3$}.
\newblock
  \url{http://math.bu.edu/people/sr/articles/ComputationsChernSimonsS2xS3_July_1_2010.pdf},
  2010.

\bibitem{maeda2016geometry}
Y.~Maeda, S.~Rosenberg, and F.~Torres-Ardila.
\newblock {The geometry of loop spaces II: Characteristic classes}.
\newblock {\em Advances in Mathematics}, 287:485--518, 2016.

\bibitem{mccarthy1978history}
J.~McCarthy.
\newblock History of {LISP}.
\newblock In {\em History of Programming Languages Conference (HOPL-I)}, pages
  173--185, 1978.

\bibitem{muranushi2016automatic}
T.~Muranushi et~al.
\newblock Automatic generation of efficient codes from mathematical
  descriptions of stencil computation.
\newblock In {\em Proceedings of the 5th International Workshop on Functional
  High-Performance Computing}, pages 17--22. ACM, 2016.

\bibitem{ollivier2011visual}
Y.~Ollivier.
\newblock {A visual introduction to Riemannian curvatures and some discrete
  generalizations}.
\newblock {\em Analysis and Geometry of Metric Measure Spaces: Lecture Notes of
  the 50th S{\'e}minaire de Math{\'e}matiques Sup{\'e}rieures (SMS),
  Montr{\'e}al}, pages 197--219, 2011.

\bibitem{DBLP:books/daglib/0005958}
B.~C. Pierce.
\newblock {\em {Types and Programming Languages}}.
\newblock {MIT} Press, 2002.

\bibitem{ricci1900methodes}
M.~Ricci and T.~Levi-Civita.
\newblock M{\'e}thodes de calcul diff{\'e}rentiel absolu et leurs applications.
\newblock {\em Mathematische Annalen}, 54(1-2):125--201, 1900.

\bibitem{minHodge}
M.~Ru.
\newblock Hodge theory on {R}iemannian manifolds.
\newblock \url{https://www.math.uh.edu/~minru/Riemann08/hodgetheory.pdf}, 2008.

\bibitem{schutz1980geometrical}
B.~F. Schutz.
\newblock {\em Geometrical methods of mathematical physics}.
\newblock Cambridge university press, 1980.

\bibitem{solomonik2015sparse}
E.~Solomonik and T.~Hoefler.
\newblock Sparse tensor algebra as a parallel programming model.
\newblock {\em Preprint at arXiv:1512.00066}, 2015.

\bibitem{steele1996evolution}
G.~L. Steele and R.~P. Gabriel.
\newblock The evolution of {Lisp}.
\newblock In {\em History of Programming Languages Conference (HOPL-II)}, pages
  231--270, 1993.

\bibitem{sussman2013functional}
G.~J. Sussman and J.~Wisdom.
\newblock {\em Functional differential geometry}.
\newblock MIT Press, 2013.

\bibitem{toth2005tensor}
V.~Toth.
\newblock {Tensor manipulation in GPL Maxima}.
\newblock {\em Preprint at cs/0503073}, 2005.

\bibitem{DBLP:journals/corr/abs-1802-04730}
N.~Vasilache, O.~Zinenko, T.~Theodoridis, P.~Goyal, Z.~DeVito, W.~S. Moses,
  S.~Verdoolaege, A.~Adams, and A.~Cohen.
\newblock Tensor comprehensions: Framework-agnostic high-performance machine
  learning abstractions.
\newblock {\em Preprint at arXiv:1802.04730}, 2018.

\end{thebibliography}

\end{document}